\documentclass[letterpaper]{article} 
\usepackage{aaai25}  
\usepackage{times}  
\usepackage{helvet}  
\usepackage{courier}  
\usepackage[hyphens]{url}  
\usepackage{graphicx} 
\def\UrlFont{\rm}  
\usepackage{natbib}  
\usepackage{caption} 
\usepackage{algorithm}
\usepackage{algorithmic}
\usepackage{newfloat}
\usepackage{listings}

\usepackage{amssymb} 
\usepackage{bm}
\usepackage{bbding} 
\usepackage{graphicx} 
\usepackage{multirow} 
\usepackage{color} 

\usepackage{hhline}
\usepackage{booktabs} 
\usepackage{boldline}
\usepackage{amsmath} 
\usepackage{array} 
\usepackage{enumerate} 
\usepackage{pdfpages}
\usepackage{subfigure}
\usepackage{tabularx}
\usepackage{makecell}

\DeclareCaptionStyle{ruled}{labelfont=normalfont,labelsep=colon,strut=off} 
\floatstyle{ruled}
\newfloat{listing}{tb}{lst}{}
\floatname{listing}{Listing}
\title{\bm{$M^3EL$}: A Multi-task Multi-topic Dataset for Multi-modal Entity Linking}

\author{
    Fang Wang\textsuperscript{\rm 1},
    Shenglin Yin\textsuperscript{\rm 1},
    Xiaoying Bai\textsuperscript{\rm 2},
    Minghao Hu\textsuperscript{\rm 2},
    Tianwei Yan\textsuperscript{\rm 3},
    Yi Liang\textsuperscript{\rm 4},
}
\affiliations{
    \textsuperscript{\rm 1}Peking University, School of Computer Science\\
    \textsuperscript{\rm 2}Advanced Institute of Big Data, Beijing\\
    \textsuperscript{\rm 3}National University of Defense Technology, College of Computer\\
    \textsuperscript{\rm 4}Xinjiang University, School of Computer Science and Technology\\
    \{fangwang, yinsl\}@stu.pku.edu.cn, baixy@aibd.ac.cn, huminghao16@gmail.com, augusyan@hotmail.com, 1049809183@qq.com
}

\begin{document}

\maketitle

\begin{abstract}


Multi-modal Entity Linking (MEL) is a fundamental component for various downstream tasks. However, existing MEL datasets suffer from small scale, scarcity of topic types and limited coverage of tasks, making them incapable of effectively enhancing the entity linking capabilities of multi-modal models. To address these obstacles, we propose a dataset construction pipeline and publish $M^3EL$, a large-scale dataset for MEL. $M^3EL$ includes 79,625 instances, covering 9 diverse multi-modal tasks, and 5 different topics. In addition, to further improve the model's adaptability to multi-modal tasks, We propose a modality-augmented training strategy. Utilizing $M^3EL$ as a corpus, train the $\textit{CLIP}_{\textit{ND}}$ model based on \textit{CLIP} (\textit{ViT}-\textit{B}-\textit{32}), and conduct a comparative analysis with an existing multi-modal baselines. Experimental results show that the existing models perform far below expectations (ACC of 49.4\%-75.8\%), After analysis, it was obtained that small dataset sizes, insufficient modality task coverage, and limited topic diversity resulted in poor generalisation of multi-modal models. Our dataset effectively addresses these issues, and the $\textit{CLIP}_{\textit{ND}}$ model fine-tuned with $M^3EL$ shows a significant improvement in accuracy, with an average improvement of 9.3\% to 25\% across various tasks. Our dataset is available at \url{https://anonymous.4open.science/r/M3EL}.

\end{abstract}
\section{Introduction}
Multi-modal Entity Linking is a crucial research direction in Natural Language Processing (NLP) and Computer Vision (CV)~\cite{sevgili2022neural}. It aims to achieve cross-modal information integration and understanding by associating entities in different modalities (e.g., text, images, etc.), and to improve the accuracy of linking to entities in the Knowledge Graph (KG). It is an important foundation for NLP downstream tasks, e.g., Information Retrieval~\cite{chang2006survey, martinez2020information} and Question-Answer systems~\cite{allam2012question, molla2006named}. Despite recent research progress, existing MEL methods face the following challenges existing in training data, thus far from meeting the requirements of real-world applications.

There are many limitations of current MEL datasets, which are summarised in Table~\ref{table:existing_datasets}. Firstly, existing datasets are generally small in scale; for instance, WIKIPerson~\cite{sun2022visual} involves 13K entities and WIKIDiverse~\cite{wang2022wikidiverse} covers 40K entities, which are insufficient for complex tasks. Secondly, the scope of modal tasks is restricted~\cite{hoffart-etal-2011-robust,moon-etal-2018-multimodal-named,gan2021multimodal}, primarily to \textit{Text}-\textit{Text}, \textit{Image}-\textit{Text}, or \textit{Image}-(\textit{Image}+\textit{Text}), while neglecting other potentially valuable tasks such as \textit{Text}-\textit{Image} and \textit{Text}-(\textit{Image}+\textit{Text}). Additionally, most datasets are confined to single topic~\cite{guo2018robust,zhang2021attention}, such as person or news, and lack the ability to generalise across topics. These constraints significantly impede the development of MEL tasks.

\begin{table*}[!t]
\scriptsize
\centering
\begin{tabular}{c|l|cccc}
\toprule
\textbf{Task} &  \textbf{Dataset} & \textbf{Source} & \textbf{Topic} & \textbf{Size} & \textbf{Modality}  \\ \midrule

 & AIDA~\cite{hoffart-etal-2011-robust} & Wikipedia & Sports & 1K docs & $T_m\rightarrow T_e$ \\
 & MSNBC\cite{cucerzan-2007-large} & Wikipedia & News & 20 docs & $T_m\rightarrow T_e$ \\
 & AQUA\cite{milne2008learning} & Wikipedia & News & 50 docs & $T_m\rightarrow T_e$ \\
 & ACE2004\cite{ratinov-etal-2011-local} & Wikipedia & News & 57 docs & $T_m\rightarrow T_e$ \\
 & CWEB\cite{guo2018robust} & Wikipedia & Web & 320 docs & $T_m\rightarrow T_e$ \\
 & WIKI\cite{guo2018robust} & Wikipedia & Common & 320 docs & $T_m\rightarrow T_e$ \\
\multirow{-7}{*}{\textbf{EL}} & Zeshel\cite{logeswaran-etal-2019-zero} & Wikia & Common & - & $T_m\rightarrow T_e$ \\ \midrule
 & Snap\cite{moon-etal-2018-multimodal-named} & Freebase & Social Media & 12K captions & $T_m,I_m \rightarrow T_e$  \\
 & Twitter\cite{adjali-etal-2020-building} & Twitter users & Social Media & 4M tweets & $T_m,I_m\rightarrow T_e,I_e$ \\
 & Movie\cite{gan2021multimodal} & Wikipedia & Movie Reviews & 1K reviews & $T_m,I_m\rightarrow T_e,I_e$ \\
 & Weibo\cite{zhang2021attention} & Baidu Baike & Social Media & 25K posts & $T_m,I_m\rightarrow T_e,I_e$ \\ 
 & WIKIDiverse\cite{wang2022wikidiverse} & Wikipedia & News & 8K captions & $T_m,I_m\rightarrow T_e,I_e$ \\ 
 & WIKIPerson\cite{sun2022visual} & Wikipedia & Person & 50K images & $T_m,I_m\rightarrow T_e,I_e$ \\ \cmidrule{2-6}
\multirow{-7}{*}{\textbf{MEL}} & \textsc{$M^3EL$} & \makecell{Wikipedia, Wikidata, \\Dbpedia, Goodreads} & \makecell{Sports, Movies, Books, \\Person, Common} & \makecell{79K instances \\ 318.5K images} & $T_m,I_m,(I+T)_m\rightarrow T_e,I_e,(I+T)_e$ \\ \bottomrule
\end{tabular}
\caption{Overview of EL and MEL datasets. $T_m$ ($T_e$) and $I_m$ ($I_e$) are abbreviations for \textit{$Text_m$} (\textit{$Text_e$}) and \textit{$Image_m$} (\textit{$Image_e$}), which represent the textual and visual contexts of mentions $m$ (or entities $e$) respectively.}

\label{table:existing_datasets}
\end{table*}

In this paper, we address these challenges by releasing the large-scale multi-task, multi-topic, multi-modal entity linking dataset, termed $M^3EL$, which has three unique key properties. First, the dataset contains 79K instances and 318.5K images, which is nearly 10 times the instances in WIKIPerson~\cite{sun2022visual} and 6.37 times the images in WIKIDiverse~\cite{wang2022wikidiverse}. Second, $M^3EL$ covers the widest range of multi-modal tasks, including: \textit{Text}-\textit{Text}, \textit{Text}-\textit{Image}, \textit{Text}-(\textit{Image}+\textit{Text}), \textit{Image}-\textit{Text}, \textit{Image}-\textit{Image}, \textit{Image}-(\textit{Image}+\textit{Text}), (\textit{Image}+\textit{Text})-\textit{Text}, (\textit{Image}+\textit{Text})-\textit{Image}, (\textit{Image}+\textit{Text})-(\textit{Image}+\textit{Text}). Finally, the $M^3EL$ dataset provides more extensive topics (movies, common, person, books, sports), while others provide only a single topic. This implies that our dataset is a more challenging setting.

Furthermore, we benchmark many existing models in our dataset, the experimental results reveal that existing models exhibit sub-optimal performance, with accuracy (ACC) ranging from 49.4\% to 75.8\%. To improve the adaptability of MEL methods, we introduce a modality-augmented training strategy to fine-tune the \textit{CLIP} (\textit{ViT}-\textit{B}-\textit{32})~\cite{clip} model. This strategy, by expanding the expressiveness of modal features, augments the training process, resulting in the model $\textit{CLIP}_{\textit{ND}}$. Following experimentation and analysis, the $\textit{CLIP}_{\textit{ND}}$ model obtained after training demonstrates significant performance improvements in ACC, with an average increase of 9.3\% to 25\%. This illustrates the considerable potential of the modal enhancement strategy in handling multi-modal data and emphasizes the research value of the $M^3EL$ dataset.

To summarize, the main contributions of this work are as follows:
\begin{itemize}
\item[1.] We present a large-scale manually labelled high-quality dataset, $M^3EL$, which covers a diverse range of topics and supports various multi-modal tasks.
\item[2.] We propose a modality-augmented training strategy. Based on this strategy, we use the $M^3EL$ dataset as a corpus to train a \textit{CLIP} (\textit{ViT}-\textit{B}-\textit{32}) model, resulting in $\textit{CLIP}_{\textit{ND}}$, which further improves the model's performance in multi-modal entity linking.
\item[3.] Experimental results indicate that our proposed large-scale, multi-task, multi-topic, multi-modal dataset serves as a high-quality pre-training corpus. While effectively enhancing the model's generalization performance, our dataset still presents a challenging benchmark.
\end{itemize}

\section{Related Work}

\noindent \textbf{Textual Entity Linking \quad}
Entity Linking (EL) is a fundamental and classical task in Natural Language Processing (NLP) that has been extensively studied. Early research introduced a variety of datasets to evaluate the performance of EL models, including high-quality manually annotated datasets such as AIDA~\cite{hoffart-etal-2011-robust}, and large-scale automatically annotated datasets such as CWEB~\cite{guo2018robust}. In addition, zero-shot entity linking datasets such as Zeshel~\cite{logeswaran-etal-2019-zero} have also been used for relevant studies. However, the performance of many EL methods on traditional datasets such as AIDA-test, MSNBC~\cite{cucerzan-2007-large}, and AQUAINT~\cite{milne2008learning} has stabilised in recent years, approaching the upper limit of performance for the task, which may indicate that performance on these datasets is nearing saturation. With the development of large-scale pre-trained language models, the latest deep learning methods have achieved over 90\% accuracy on the AIDA dataset, which also implies that current methods may be approaching a performance bottleneck~\cite{de2020autoregressive}. Consequently, to advance the field, researchers have designed more challenging tasks such as zero-shot entity linking~\cite{logeswaran-etal-2019-zero,wu2019scalable}, global coherence exploitation~\cite{chen2020improving}, NIL prediction~\cite{rao2013entity},  and end-to-end solutions for emerging entities~\cite{broscheit2020investigating}. These tasks are intended to drive further development and breakthroughs in the field of EL.

\noindent \textbf{Multi-modal Entity Linking \quad}
In recent years, the task of MEL has gained significant attention in the field of NLP. This task aims to utilize both textual and visual information to map ambiguous mentions to entities in a Knowledge Graph (also known as Knowledge Base). Moon et al. first demonstrated the effectiveness of image information in dealing with ambiguous and short textual mentions in social media posts, proposing a zero-shot framework to integrate textual, visual and lexical information for entity linking~\cite{moon-etal-2018-multimodal-named}. However, their proposed dataset is unavailable due to GDPR regulations. Adjali et al. developed a framework for automatically constructing an MEL dataset from Twitter, with small data scale, scarcity of entity types, and ambiguous mentions restricting the usefulness of the dataset~\cite{adjali2020building}. Zhang et al. construct a Chinese multi-source social media multi-modal dataset based on the social media platform Weibo, focusing solely on person entities~\cite{zhang2021attention}. Gan et al. published a MEL dataset focusing on roles and characters in the movie domain based on movie reviews~\cite{gan2021multimodal}. Both of these two datasets have a single entity type and the generalisation of the training model is weak. Wang et al. constructed a multi-modal entity linking dataset with diverse contextual themes and entity types, including multiple modal tasks~\cite{wang2022wikidiverse}. However, all those works still face issues such as small data size, limited entity types, and incomplete multi-modal tasks, therefore, the diversity of entity types and modal tasks in datasets needs further exploration and optimization.

\begin{figure*}[!h]
    \centering
    \includegraphics[width=1\textwidth]{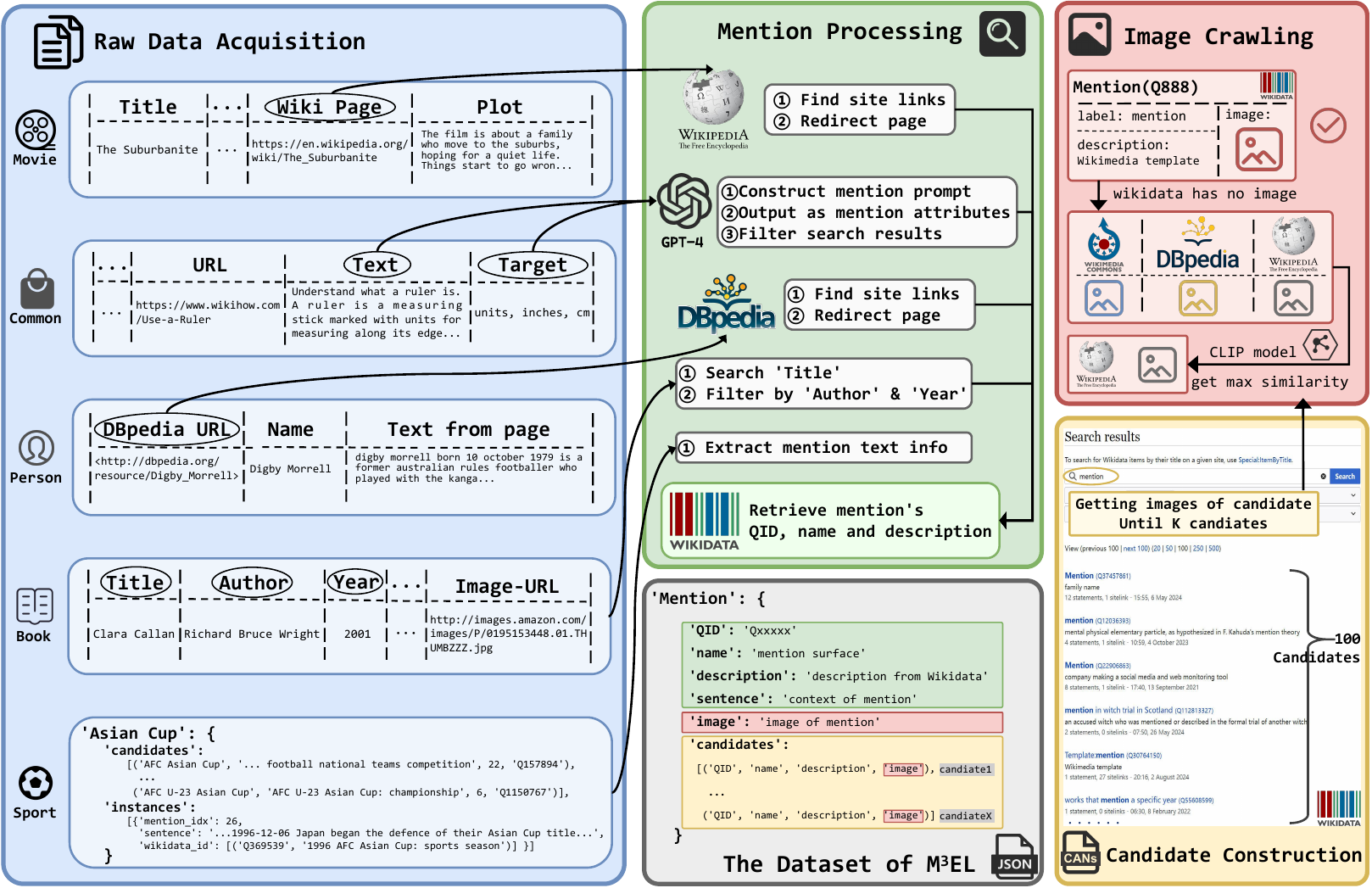}
    \caption{Overview of the $M^3EL$ data construction pipeline.}
    \label{fig:overview}
\end{figure*}

\noindent \textbf{Entity Linking Dataset \quad}
Our $M^3EL$ dataset also encompasses multi-modal image-text datasets. While existing related studies (~\cite{biten2019good,tran2020transform,liu2020visual,zhang2018adaptive,lu2018visual}) have proposed larger-scale image-text datasets, the entities in these datasets lack detailed explanations and are not linked to a unified knowledge graph. The Flickr 30k~\cite{young2014image} and MSCOCO~\cite{chen2015microsoft} datasets made strides in size and annotate entities with descriptive sentences. However, the mention information in these sentences tends to be more ambiguous. Although the WIKIPerson~\cite{sun2022visual} and WIKIDiverse~\cite{wang2022wikidiverse} datasets address some of these issues, they contain only person entities and focus mainly on specific multi-modal tasks, such as \textit{Text-Text} and \textit{Image-Image}. The comprehensiveness of the task types still needs improvement.
\section{Problem Formulation}
Multi-modal entity linking maps mentions in multi-modal contexts to corresponding entities in a target Knowledge Graph (\textit{KG}). Since the large number of entities in \textit{KG}, traversing all entities for linking significantly increases time costs, thus candidate sets of entities are usually predefined. On this basis, we construct a candidate set for each mention, which may or may not contain the correct corresponding entity, accurately reflecting the challenges inherent in multi-modal entity linking tasks.

Formally, let \( E \) represent the set of entities in the \textit{KG}, typically comprising millions of entities. Let \( C \) denote the candidate set in the \textit{KG} for each mention \( m \), where \( C \subseteq E \). Each mention \( m \) or entity \( e \in C \) is associated with its corresponding visual context \( V_m \), \( V_e \), and textual context \( T_m \), \( T_e \). Here, \( T_m \) and \( T_e \) denote the textual context surrounding \( m \) and \( e \), respectively. \( V_m \) refers to the images related to \( m \), while \( V_e \) refers to the images associated with \( e \) in the \textit{KG}. Therefore, the multi-modal task of identifying the corresponding entity for mention \( m \) can be defined as follows: 
\vspace{-0.18cm}
\begin{equation}
    Sim(M_m,M_e) = \mathop{\arg\max} \limits_{e_i \in C}\Psi(En(M_m), En(M_e)),
    \label{eq:1}
\end{equation}
where $M_m$ and $M_e$ represents the modal information of the target mention $m$ and the candidate entity $e$, respectively, as detailed in Table~\ref{table:different_modal_combine}. $En$ denotes \textit{Encoder}. The function $\Psi$ denotes the similarity score between the $m$ and the $e$.

\begin{table}[t]
\centering
\resizebox{\columnwidth}{!}{%
\begin{tabular}{c|c|c}
\toprule
$\bm{M_m}$ & $\bm{M_e}$ & \textbf{Multi-modal Tasks} \\ \midrule
$T_m$ & $T_e$ & $T_m - T_e \ \vert\  T_m - I_e  \ \vert\  T_m - (I+T)_e$ \\ \midrule
$I_m$ & $I_e$ & $I_m - T_e  \ \vert\   I_m - I_e  \ \vert\  I_m - (I+T)_e$ \\ \midrule
$(I+T)_m$ & $(I+T)_e$ & $(I+T)_m - T_e \ \vert\  (I+T)_m - I_e  \ \vert\  (I+T)_m - (I+T)_e$ \\ \bottomrule
\end{tabular}%
}
\caption{Introduction to different modal information and modal tasks. $T_m$ ($T_e$) denotes the textual information of the mention (entity), while $I_m$ ($I_e$) represents the visual information of the mention. $(I+T)_{m(e)}$ refers to the combined textual and visual information of the mention (entity).}
\label{table:different_modal_combine}
\end{table}

\section{Data Setups for $\boldsymbol{M^3EL}$}


The framework of the data construction pipeline of $M^3EL$ is illustrated in Figure 1. In the following, we detail the construction of $M^3EL$.



\subsection{Data Collection}
\noindent \textbf{Raw data Acquisition. \quad}To construct the foundational dataset for $M^3EL$, we collected raw data from Kaggle\footnote{\url{https://www.kaggle.com/}} that includes URL information related to DBpedia\footnote{\url{https://www.dbpedia.org/}}, Wikipedia\footnote{\url{https://www.wikipedia.org/}}, or Wikidata\footnote{\url{https://www.wikidata.org/wiki/Wikidata:Main_Page}}. These data files are diverse in format, including JSON, CSV, Pickle, and more. Since DBpedia, Wikipedia and Wikidata pages can interlink, where Wikidata's QID attributes can be used as entity labels, these datasets are well-suited as raw data sources for research in multi-modal entity linking. In addition, these Kaggle datasets are derived from real-world data science scenarios and cover a wide range of topics, reflecting the complexity and diversity of data in the real world. Specifically, we utilized \textit{the Wikipedia Movie Plots}~\footnote{\url{https://www.kaggle.com/datasets/jrobischon/wikipedia-movie-plots?select=wiki_movie_plots_deduped.csv}} dataset as the raw data for the movies topic, the \textit{Wiki-Wiki}~\footnote{\url{https://drive.google.com/file/d/1Oebe1sbbixX7FWHX813diCdqReh1IyWH/view}} dataset for common knowledge topic, the \textit{People Wikipedia Data}\footnote{\url{https://www.kaggle.com/datasets/sameersmahajan/people-wikipedia-data}} for person topic, the \textit{Books Datasets}~\footnote{\url{https://www.kaggle.com/datasets/saurabhbagchi/books-dataset}} for book topic, and the \textit{AIDA-CoNLL-Test}\footnote{\url{https://www.mpi-inf.mpg.de/departments/databases-and-information-systems/research/ambiverse-nlu/aida/downloads}} for the sports topic.

\noindent \textbf{Data Filter and clean. \quad}For the raw dataset, we performed the following data cleaning steps: 1) Removed non-English expressions and single-character mentions. 2) Deleted mentions without associated images or whose images could not be downloaded. Following these pre-processing steps, we finally obtained a total of 82K mentions, each comprising a text-image pair, which serves as the base dataset for $M^3EL$.

\subsection{Data Curation}
\noindent \textbf{Mention Processing. \quad} 
Due to the varying formats of raw data across different topics, we describe in detail the process of obtaining relevant mentions for each topic separately.

For the raw data on the movie topic, we utilized the "Title" attribute from the \textit{wiki\_movie\_plots\_deduped.csv} file to extract movie titles as mentions. The "Plot" attribute was employed to provide contextual information related to these mentions. Additionally, we accessed the mention-related Wikipedia pages through the URLs provided in the "Wiki Page" attribute. From these Wikipedia pages, we further navigated to the corresponding Wikidata pages to obtain fundamental information, including the QID and descriptions.

For the raw data on commons topic, we utilized the instances provided by the "Target" attribute in the \textit{wikihow.csv} file as mentions, and paragraphs provided by the "Text" attribute as contextual information for mentions. By constructing a mention semantic understanding prompt, we input it into the large language model GPT-4 to obtain semantic information related to the mentions. Subsequently, each mention was searched in Wikidata, and the Wikidata entry that best matches the mention is filtered based on the semantic information obtained from GPT-4, and relevant information such as QID and description is obtained.

For the raw data on person topic, we utilized the \textit{people\_wiki.csv} file, specifically the "name" attribute, as the reference for mentions, and the "text" attribute, which provides the context for these mentions. The associated DBpedia pages were accessed via the "URL" attribute. Subsequently, we navigated to the corresponding Wikidata pages to obtain foundational information, including the QID and descriptions.

For the raw data on book topics, we utilized the "Book-Title" attribute of the \textit{books.csv} file as a mention for searching the corresponding Wikidata pages to retrieve basic information, including the QID and description. To ensure the accuracy of the search results, we cross-validated them using the Book-Author, Year of Publication, and Publisher information provided in the file. Mentions that could not be found on the Wikidata page were subsequently removed.

For the raw data on sport topic, the \textit{AIDA\_B\_dict\_} \textit{raw.pickle} file provides mentions, contextual information, and associated QID information in Wikidata in the form of a dictionary. We directly utilized the textual information from the raw data without any processing.

\noindent \textbf{Candidate Construction. \quad} 
In most entity linking tasks, Wikidata is used as the target Knowledge Graph, linking mentions to their corresponding entities. However, due to the large scale of Wikidata, which contains approximately 95 million entries, matching every mention by traversing all entities would consume substantial computational resources and time. Therefore, to enhance the quality and usability of the dataset, we employ an optimization strategy: for each mention, a refined subset is filtered from Wikidata to serve as the candidate set. This set contains both possible and incorrect corresponding candidates, which truly reflects the challenges of the multi-modal entity linking task.

The specific steps are as follows: 1) We search for the mention surface on the Wikidata page and retrieve the top 100 relevant search results. For each search result, we access its Wikidata detail page and systematically extract the key information, including the entity's QID, entity name, and description, among other textual data. 2) We enter the stage of obtaining candidate entity images, and when the number of candidates satisfying complete textual and visual information reaches a predetermined threshold of $k$, the candidate set construction is completed. This process ensures the completeness of the information and the high quality of the dataset, which provides a solid foundation for the subsequent data analysis and model training.

\noindent \textbf{Image Crawling. \quad} 
After obtaining the mention-related textual information of different topics, in order to obtain the image information corresponding to the mentions, we developed a comprehensive image crawling frameworks based on Wikidata, Wikipedia and Wikimedia\footnote{\url{https://commons.wikimedia.org/wiki/Main_Page}} pages. Furthermore, we utilized a pre-trained \textit{CLIP} (\textit{ViT}-\textit{B}-\textit{32}) model to effectively filter the most relevant images by leveraging the joint representation of visual and textual information. Especially for the book topic, we directly use the URLs of book covers provided in the raw data to obtain relevant images. For invalid URLs, we implemented a supplementary crawler that automatically searches for and retrieves relevant images using book titles as keywords on Goodreads\footnote{\url{https://www.goodreads.com}} and Amazon\footnote{\url{https://www.amazon.com}}. These steps ensure that we are able to obtain the appropriate image information for each mention, enriching the multi-modal characteristics of the dataset.

\subsection{Data Analysis}
\begin{figure}[t]
    \centering
    \subfigure[\scriptsize{Entity topic distribution.}]{\includegraphics[width=0.19\textwidth]{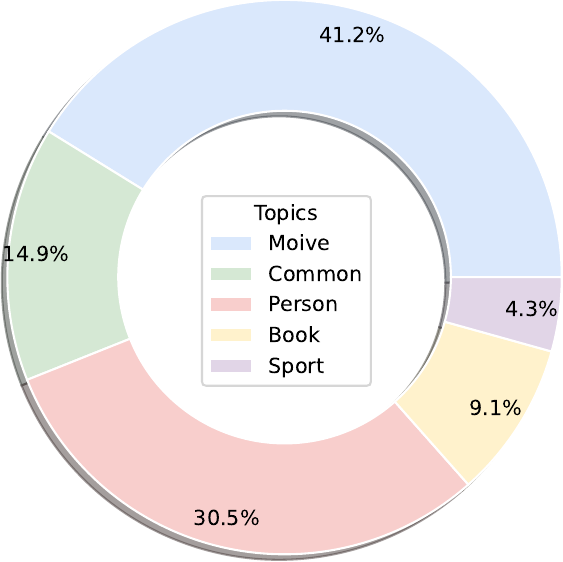}}
    \hspace{0.1in}
    \subfigure[\tiny{Different type candiates distribution.}]{\includegraphics[width=0.25\textwidth]{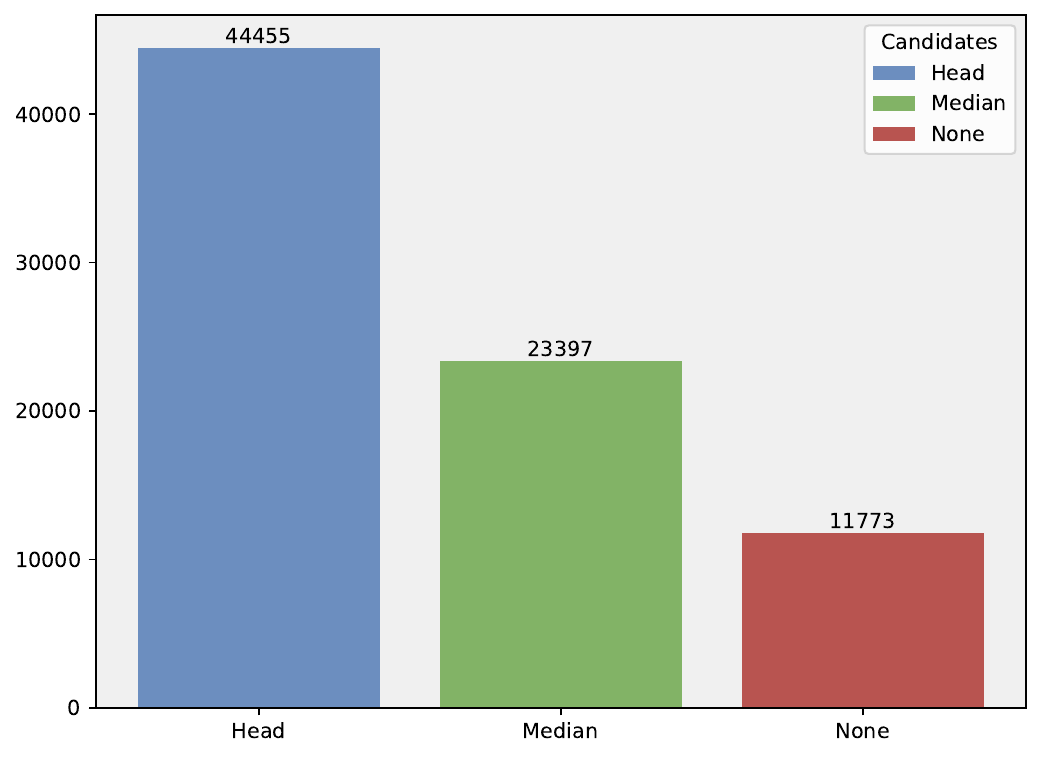}}
    \caption{More statistics of $M^3EL$. (a) Entity topic type distribution. (b) Distribution of correct entity positions within the candidates.}
    \label{figure: Statistics_of_type}
\end{figure}

\noindent \textbf{Size and Difficulty Measure. \quad}
The statistics of $M^3EL$ dataset in detail are shown in Table~\ref{table:Data_Analysis}. This dataset was constructed from 66,342 articles and contains a total of 79K entities, covering 318.5K images. Each entity is mapped to a specific entity in Wikidata. Notably, many entities appear multiple times in different sentences of the articles, ensuring that the entities can be fully learned. In addition, Figure~\ref{figure: Statistics_of_type}(a) reports the distribution of entity topics. Unlike existing MEL datasets, which are typically constructed for specific scenarios or tasks, $M^3EL$ covers 5 distinct topics and involves 9 types of multi-modal linking tasks. 

Firstly, we compare the surface form similarity between mentions and ground-truth entities. It is observed that 41.3\% of the mentions differ from the surface forms of the ground-truth entities. This significant discrepancy in surface forms presents a challenge for multi-modal entity linking tasks. Secondly, we report the candidates for each mention in Figure~\ref{figure: Statistics_of_type}(b). We observe the following: 1) 99.8\% of mentions have three candidates, with 55.8\% of these mentions having the correct entity ranked first in the candidate set, and 29.4\% having the correct entity ranked elsewhere within the candidate set. 2) 14.8\% of mentions have candidate sets that do not include the correct entity. This indicates that the candidate set reflects the real-world scenarios of entity linking to a certain extent.

\begin{table}[t]
\centering
\resizebox{\columnwidth}{!}{%
\begin{tabular}{c|c|c|c|c|c}
\toprule
\textbf{Topic} & \textbf{Documents} & \textbf{Mention} & \textbf{Candidate} & \textbf{Image} & \textbf{Tasks} \\ \midrule
Movie & 32131 & 32800 & 98400 & 131200 & \multirow{7}{*}{9} \\ \cmidrule{1-5}
Common & 1699 & 11826 & 35478 & 47304 &  \\ \cmidrule{1-5}
Person & 26873 & 24278 & 72834 & 97112 &  \\ \cmidrule{1-5}
Book & 5408 & 7276 & 21828 & 29104 &  \\ \cmidrule{1-5}
Sport & 231 & 3445 & 10335 & 13780 &  \\ \midrule
ALL & 66342 & 79625 & 238875 & 318500 & 9 \\ \bottomrule
\end{tabular}%
}
\caption{Statistics of $M^3EL$. Documents represent the raw textual contexts of mentions, Tasks refers to 9 different multi-modal tasks.}
\label{table:Data_Analysis}
\end{table}

\noindent \textbf{Diversity of multi-modal tasks. \quad}
Compared to existing datasets, $M^3EL$ offers a more comprehensive range of multi-modal tasks. Traditional multi-modal entity linking datasets, such as Snap~\cite{moon-etal-2018-multimodal-named}, Twitter~\cite{adjali-etal-2020-building}, and Weibo~\cite{zhang2021attention}, are designed for specific tasks. Although datasets like WIKIPerson~\cite{sun2022visual} and WIKIDiverse~\cite{wang2022wikidiverse} introduce various forms of multi-modal entity linking tasks, including $I_m$-$T_e$, $I_m$-$I_e$, and $(I$+$T)_m$-$I_e$, they ignored other potential linking tasks. In contrast, $M^3EL$ significantly expands the scope of tasks, covering 9 distinct types, specifically, $T_m$-$T_e$, $T_m$-$I_e$, $T_m$-($I$+$T)_e$, $I_m$-$T_e$, $I_m$-$I_e$, $I_m$-($I$+$T)_e$, ($I$+$T)_m$-$T_e$, ($I$+$T)_m$-$I_e$, ($I$+$T)_m$-($I$+$T)_e$. This task diversity greatly enhances MEL's ability to generalise across a wide range of multi-modal related tasks, enabling it to better accommodate the diverse needs of users in the real world.

\noindent \textbf{Coverage of topic types. \quad}
As shown in Table~\ref{table:existing_datasets}, most of the existing training datasets focus on a limited range of entity topic types, such as person and news. WIKIDiverse~\cite{wang2022wikidiverse} extends the types of topic to multiple categories by developing crawling code to extract multi-modal information related to text and images from Wikinews, making it state-of-the-art, but all categories belong to news topic. However, our constructed $M^3EL$ includes 5 types of topics, specifically, movies, common knowledge, person, books, and sports. Moreover, Figure~\ref{figure: Statistics_of_type}(a) illustrates the distribution of $M^3EL$ across different topics.

\section{Model with Augmented Strategy}

\noindent \textbf{\textit{CLIP} Definition \quad}
The Contrastive Language–Image Pre-training (\textit{CLIP}) method has demonstrated significant effectiveness in training visual models using language supervision. In each training step, a large batch of $N$ pairs of images and texts \(\{x_I, x_T\}\) is randomly sampled from the training dataset. Subsequently, data augmentation is applied to the images to simulate various real-world visual variations. The augmented images and their corresponding texts are then processed by dedicated encoders and normalization functions \(f_I\) and \(f_T\) to extract image and text features. These features are utilized to compute the InfoNCE loss, a type of contrastive loss that distinguishes between matched (positive) and mismatched (negative) image-text pairs. The training loss is represented by the following equation:

\begin{equation}
    L = \sum_{i=1}^N \log \frac{exp(sim(f_I(aug_I(x_I^i)), f_T(x_T^i))/ \tau)}{\sum_{j=1}^N exp(sim(f_I(aug_I(x_I^j)), f_T(x_T^j))/ \tau)},
    \label{eq:2}
\end{equation}
where $(x_I^i, x_T^i)$ represents the $i$-th image-text pair, and $aug_I()$ denotes the image augmentation function. The $sim(-, -)$ function measures similarity using the dot product, while the temperature parameter $\tau$ is a learnable variable used to scale the logits.

\noindent \textbf{Augmented Strategy \quad}
In Equation~\ref{eq:2}, while the standard \textit{CLIP} loss applies augmentation to images to enhance the model's adaptability and robustness to visual variations, the textual input remains constant throughout the training process. This approach ignores the potential diversity and ambiguity inherent in textual data. To address this issue, we propose a strategy to augment the textual data with entity description information, denoted as $aug_T$, where \(x_T\) is replaced by $aug_T(x_T)$ as the input to \(f_T\). By applying text augmentation, more detailed and precise textual information is input into the model, enhancing its ability to handle semantic ambiguities and ultimately improving its performance on multi-modal tasks.
\section{Experiments}

\begin{table*}[h]
\small
\centering
\begin{tabular}{ccccccccc}
\toprule
 & \textbf{\bm{$\textit{S}_1$}} & \textbf{\textit{S}} & \textbf{$\textit{M-S}_{\bm{1}}$} & \textbf{\textit{M-S}} & \textbf{$\textit{S}_{\bm{1}}\textit{-M}$} & \textbf{\textit{S-M}} & \textbf{$\textit{S}_{\bm{1}}\textit{-M-D}$} & \textbf{\textit{S-M-D}} \\ \midrule
\textbf{ALIGN} & 0.674 & 0.675 & 0.683 & 0.683 & 0.680 & 0.681 & 0.701 & 0.697 \\
\textbf{BLIP-2} & 0.738 & 0.739 & 0.745 & 0.746 & 0.743 & 0.744 & 0.759 & 0.753 \\
\textbf{CLIP} & 0.742 & 0.748 & 0.754 & 0.758 & 0.749 & 0.753 & 0.771 & 0.763 \\
\textbf{FLAVA} & 0.654 & 0.649 & 0.661 & 0.655 & 0.659 & 0.653 & 0.681 & 0.699 \\
\textbf{OWL-ViT} & 0.487 & 0.487 & 0.495 & 0.494 & 0.489 & 0.488 & 0.490 & 0.492 \\
\textbf{SigLIP} & 0.711 & 0.714 & 0.722 & 0.724 & 0.715 & 0.717 & 0.736 & 0.726 \\
\bm{$CLIP_{ND}$} & \textbf{0.885} & \textbf{0.882} & \textbf{0.899} & \textbf{0.897} &\textbf{ 0.890} &\textbf{ 0.888} & \textbf{0.902} & \textbf{0.898} \\ \bottomrule
\end{tabular}%
\caption{Experimental results of the baseline model on various $M^3EL$ versions.The $M^3EL$ dataset comes in various versions, each made up of composite text and images. \textit{M} indicates the mention's surface, short for \textit{Mention}. \textit{S} its full context, short for \textit{Sentence}. $\textit{S}_1$ a single sentence with the mention, short for $\textit{Sentence}_1$, and \textit{D} the mention's description, short for \textit{Description}. Formats include \textit{M-S} (mention: sentence), \textit{S-M} (sentence mention), and \textit{S-M-D} (sentence mention: description), among others.}
\vspace{-0.08cm}
\label{table:various_M3EL_versions}
\end{table*}

\subsection{Experimental Setups}

\noindent \textbf{Datasets \quad}
To address the limited scale, coverage of modal tasks, and scarcity of entity topics in existing datasets, we constructed a new dataset named $M^3EL$, which consists 79,625 instances. In total, we collected 318,500 images accompanied by textual captions. The $M^3EL$ dataset has been divided into training, validation, and test sets in the ratio of 8:1:1.

To further verify the quality of the constructed dataset, we also investigated other existing multi-modal datasets:

\begin{itemize}
    \item {\textbf{DiffusionDB}~\cite{wangDiffusionDBLargescalePrompt2022}} is the first large-scale text-to-image cue dataset. We chose the subset "2m\_random\_1k", which contains 1000 image-text pairs, primarily aimed at the task of $I_m$-$T_e$.
    \item {\textbf{WIKIPerson}~\cite{sun2022visual}} is a high-quality visual person linking dataset designed for Visual Named Entity Linking tasks. We selected the test set containing 6,142 images, along with their labels (Wikidata QIDs) and descriptions, primarily aimed at the task of $I_m$-$T_e$, $I_m$-$I_e$, and $I_m$-($I$+$T)_e$.
    \item {\textbf{WIKIDiverse}~\cite{wang2022wikidiverse}} is a high-quality, manually annotated MEL dataset consisting of diverse contextual topics and entity types derived from Wikinews. We selected the test set comprising 1,570 image-caption pairs, primarily designed to implement $T_m$-$T_e$, $I_m$-$I_e$, ($I$+$T)_m$-$I_e$ and ($I$+$T)_m$-($I$+$T)_e$ tasks.
\end{itemize}

\noindent \textbf{Baselines \quad}
Given the diverse range of modality tasks encompassed by our proposed multi-modal dataset, the number of comparable models is notably limited. In order to establish a valid benchmark for evaluation, we report on the performance of several state-of-the-art methods for entity linking and visual entity recognition, including \textit{ALIGN}~\cite{align}, \textit{BLIP-2}~\cite{blip2}, \textit{CLIP}~\cite{clip}, \textit{FLAVA}~\cite{flava}, \textit{OWL-ViT}~\cite{OWL-ViT}, and \textit{SigLIP}~\cite{siglip}. Additionally, we present our own optimized approach, $\textit{CLIP}_{\textit{ND}}$.

\noindent \textbf{Implementation Details \quad}
We employed the \textit{CLIP} (\textit{ViT}-\textit{B}-\textit{32}) architecture, which comprises both text and image encoders, each with a hidden state dimension of 768 and 12 multi-head attention mechanisms. We trained on an A100 GPU with 40GB of memory with 35 epochs and a batch of 256. The AdamW optimizer was applied with an initial learning rate of 1e-5, betas=(0.9, 0.98), eps=1e-6, and weight\_decay=0.001.


\noindent \textbf{Evaluation Metric \quad}
The primary metric we evaluate is the accuracy (ACC) of linking entities to KG. Accuracy is defined by the following formula:

\begin{equation}
    ACC = \frac{\sum_{i=1}^{N_{correct}} M_m-M_e}
    {\sum_{i=1}^{N} M_m-M_e},
    \label{eq:4}
\end{equation}
where $N_{correct}$ represent the number of correctly linked entity mentions, $N$ represent the total number of mentions, $M_m$-$M_e$ denotes different modalities of mention-entity linking task.

\subsection{Experimental Results}

\begin{table}[]
\centering
\resizebox{\columnwidth}{!}{%
\begin{tabular}{ccccc}
\toprule
\textbf{} & \textbf{Diffusion} & \textbf{WIKIPerson} & \textbf{WIKIDiverse} & \bm{$M^3EL$} \\ \midrule
\textbf{ALIGN} & 0.468 & \underline{0.826} & 0.707 & 0.683 \\
\textbf{BLIP-2} & 0.552 & 0.652 & 0.677 & \underline{0.746} \\
\textbf{CLIP} & 0.517 & \underline{0.873} & 0.762 & 0.758 \\
\textbf{FLAVA} & 0.582 & 0.529 & \underline{0.718} & 0.655 \\
\textbf{OWL-ViT} & 0.421 & 0.348 & \underline{0.609} & 0.494 \\
\textbf{SigLIP} & 0.539 & \underline{0.771} & 0.743 & 0.724 \\
\bm{$CLIP_{ND}$} & \textbf{0.724} & \underline{\textbf{0.916}} & \textbf{0.796} & \textbf{0.897} \\ \bottomrule
\end{tabular}%
}
\caption{Experimental results of the baseline models on each dataset. the best-performing model on each dataset is highlighted in bold, while the best performance of each model across different datasets is indicated with an underline.}
\vspace{-0.35cm}
\label{table: Experimental_results}
\end{table}

\noindent \textbf{Main Results \quad}
In Table~\ref{table:various_M3EL_versions}, we provide a comprehensive analysis of multi-modal task performance on the $M^3EL$. The experimental results are shown: Firstly, $\textit{CLIP}_{\textit{ND}}$ achieves significant performance improvements over the baseline model on different versions of the multi-modal dataset. Specifically, on the \textit{M-S} form, $\textit{CLIP}_{\textit{ND}}$ demonstrated a 13.9\% increase in accuracy compared to the baseline, with an average improvement of 7\% across other datasets. Notably, $\textit{CLIP}_{\textit{ND}}$ and \textit{CLIP} share identical parameters and computational costs during training. Secondly, when auxiliary text information (e.g., entity names and descriptions) is added to the multi-modal task, the linking accuracy of the baseline model increases significantly, by an average of 1.2\%. This suggests that the level of detail and clarity of the auxiliary text plays a key role in enhancing the model ability. Thirdly, all models consistently performed better when utilizing the $\textit{S}_1\textit{-M-D}$ format for multi-modal tasks compared to the \textit{S-M-D} format. This is mainly due to the $\textit{S}_1\textit{-M-D}$ format's ability to convey more precise information about the entities. In contrast, the \textit{S-M-D} format may truncate key entity and description information due to text length limitations, thereby negatively impacting model performance. Additionally, while most models perform better on multi-modal tasks that contain richer textual information ($S$format), for models focusing on visual modality such as \textit{FLAVA} and \textit{OWL-ViT}, more textual information may introduce noise and degrade link performance.

\noindent \textbf{Further Analysis \quad}
In Table~\ref{table: Experimental_results}, we report the accuracy of the model in detail. Based on the experimental results, we observe that our trained multi-modal approach, $\textit{CLIP}_{\textit{ND}}$, outperforms all baseline methods that rely solely on the raw modal information, especially in the dataset WIKIPerson, where the accuracy is improved by 4.3\% to 56.8\%. This significant performance boost is primarily attributed to the augmented feature strategy implemented during the training process. Secondly, except for the \textit{BLIP-2} model, other models did not achieve the highest link accuracy on the $M^3EL$. Compared to the best-performing model, these models exhibited an average accuracy deficit of 7.6\%. Lastly, the model performed the worst on the Diffusion dataset, due to the ambiguous mentions in the textual information paired with images, making linking more hard. Models performed most consistently in the WIKIDiverse, as the surface of mentions in the text-image pairs were more explicit. Most models performed best in the dataset WIKIPerson, due to the availability of bounding boxes within images where the mentions were located, allowing the models to access more specific and accurate image features. Overall, the $M^3EL$ dataset combines the strengths of various datasets by providing clear mention and diverse text-image features, which helps enhance the generalization ability of multi-modal models.

\begin{figure}[t]
    \centering    \includegraphics[width=0.9\columnwidth]{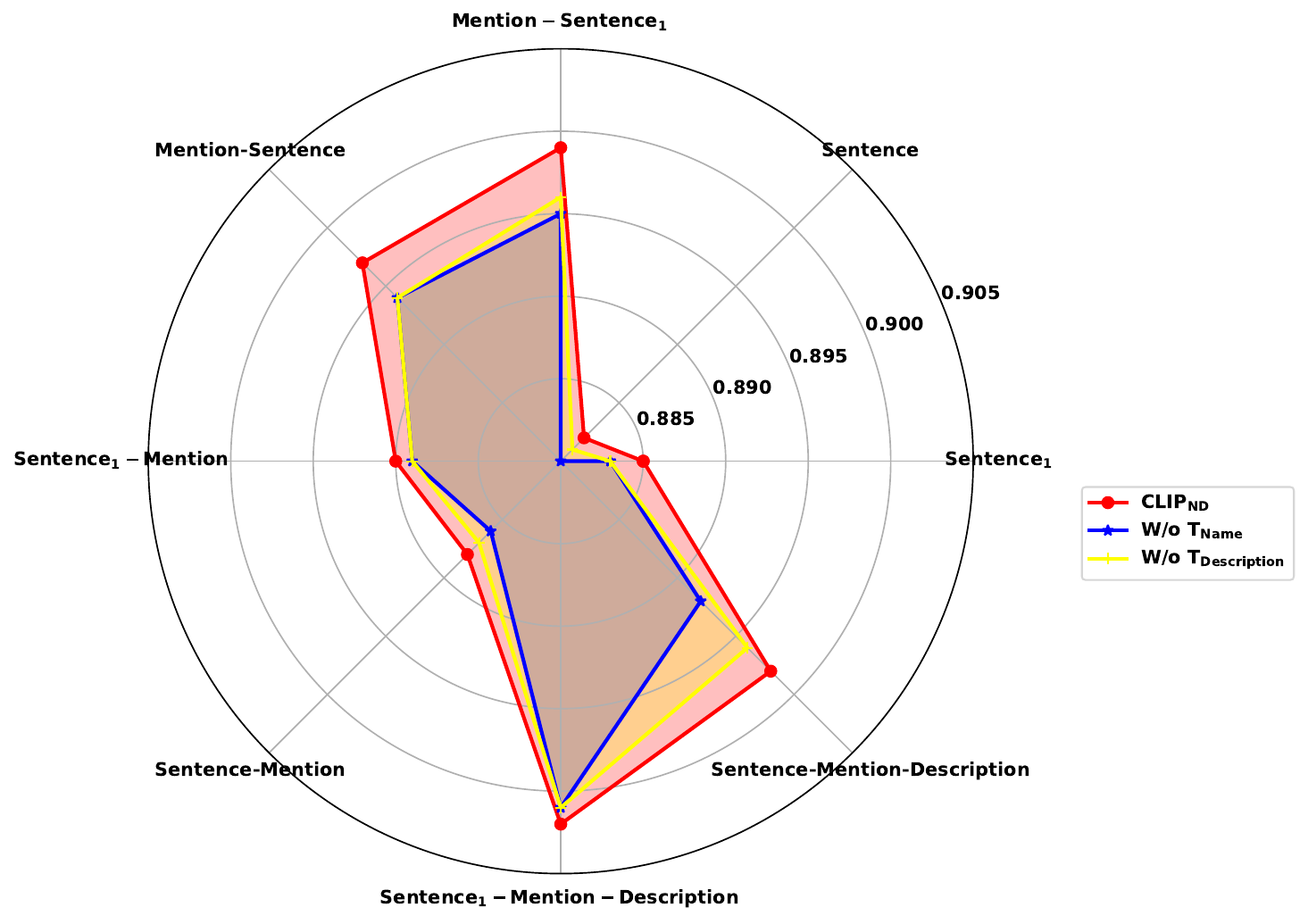}
    \caption{Ablation study to analyze modality absence of mention’s different textual information. W/o $T_{name}$ or $T_{description}$ stands for $\textit{CLIP}_{\textit{ND}}$ trained without the corresponding inputs.}
    \vspace{-0.35cm}
    \label{fig:ablation}
\end{figure}

\noindent \textbf{Ablation Analysis \quad}
We conducted an ablation study, and the experimental results are presented in Figure~\ref{fig:ablation}. It is evident that the model achieves optimal performance when it utilizes both mentions' name and description information as multi-modal context. In contrast, models trained using only a single source of information (e.g., only name or description) performed poorly. In addition, an interesting phenomenon was observed in the experiments: when the model used only sentences $\textit{S}_1$ involving mentions as context, the performance was better than using complete sentences \textit{S}. This suggests that models trained with text augmentation strategies have better comprehension in capturing semantic information related to mentions, whereas the use of longer texts may introduce noise, which reduces model performance.
\section{Conclusion}
We propose $M^3EL$, a large-scale multi-task and multi-topic dataset for multi-modal entity linking, encompassing 79,625 mentions across 5 different topics. This dataset provides detailed mention names, descriptions, contextual information, and 318.5K image-related data, covering 9 diverse multi-modal tasks. Compared to existing datasets, $M^3EL$ has advantages in terms of entity type coverage, multi-modal task diversity, and good scalability, ultimately contributing to enhancing the performance of linking models in MEL tasks. Leveraging the rich textual information of $M^3EL$, we propose augmented modality strategies to train a model $\textit{CLIP}_{\textit{ND}}$ and effectively improve its performance in MEL. In future work, we will continue to update $M^3EL$ to better support research in MEL. We will explore more complex real-world tasks, such as dynamic EL that requires complex linking in audio or video, and investigate the capabilities of $M^3EL$ in large language models.

\bibliography{aaai25}

\begin{thebibliography}{37}
\providecommand{\natexlab}[1]{#1}

\bibitem[{Adjali et~al.(2020{\natexlab{a}})Adjali, Besan{\c{c}}on, Ferret, Le~Borgne, and Grau}]{adjali-etal-2020-building}
Adjali, O.; Besan{\c{c}}on, R.; Ferret, O.; Le~Borgne, H.; and Grau, B. 2020{\natexlab{a}}.
\newblock Building a Multimodal Entity Linking Dataset From Tweets.
\newblock In \emph{Proceedings of the 12th Language Resources and Evaluation Conference}, 4285--4292. Marseille, France: European Language Resources Association.
\newblock ISBN 979-10-95546-34-4.

\bibitem[{Adjali et~al.(2020{\natexlab{b}})Adjali, Besan{\c{c}}on, Ferret, Le~Borgne, and Grau}]{adjali2020building}
Adjali, O.; Besan{\c{c}}on, R.; Ferret, O.; Le~Borgne, H.; and Grau, B. 2020{\natexlab{b}}.
\newblock Building a multimodal entity linking dataset from tweets.
\newblock In \emph{Proceedings of the Twelfth Language Resources and Evaluation Conference}, 4285--4292.

\bibitem[{Allam and Haggag(2012)}]{allam2012question}
Allam, A. M.~N.; and Haggag, M.~H. 2012.
\newblock The question answering systems: A survey.
\newblock \emph{International Journal of Research and Reviews in Information Sciences (IJRRIS)}, 2(3).

\bibitem[{Biten et~al.(2019)Biten, Gomez, Rusinol, and Karatzas}]{biten2019good}
Biten, A.~F.; Gomez, L.; Rusinol, M.; and Karatzas, D. 2019.
\newblock Good news, everyone! context driven entity-aware captioning for news images.
\newblock In \emph{Proceedings of the IEEE/CVF conference on computer vision and pattern recognition}, 12466--12475.

\bibitem[{Broscheit(2020)}]{broscheit2020investigating}
Broscheit, S. 2020.
\newblock Investigating entity knowledge in BERT with simple neural end-to-end entity linking.
\newblock \emph{arXiv preprint arXiv:2003.05473}.

\bibitem[{Chang et~al.(2006)Chang, Kayed, Girgis, and Shaalan}]{chang2006survey}
Chang, C.-H.; Kayed, M.; Girgis, M.~R.; and Shaalan, K.~F. 2006.
\newblock A survey of web information extraction systems.
\newblock \emph{IEEE transactions on knowledge and data engineering}, 18(10): 1411--1428.

\bibitem[{Chen et~al.(2020)Chen, Wang, Jiang, and Lin}]{chen2020improving}
Chen, S.; Wang, J.; Jiang, F.; and Lin, C.-Y. 2020.
\newblock Improving entity linking by modeling latent entity type information.
\newblock In \emph{Proceedings of the AAAI conference on artificial intelligence}, volume~34, 7529--7537.

\bibitem[{Chen et~al.(2015)Chen, Fang, Lin, Vedantam, Gupta, Doll{\'a}r, and Zitnick}]{chen2015microsoft}
Chen, X.; Fang, H.; Lin, T.-Y.; Vedantam, R.; Gupta, S.; Doll{\'a}r, P.; and Zitnick, C.~L. 2015.
\newblock Microsoft coco captions: Data collection and evaluation server.
\newblock \emph{arXiv preprint arXiv:1504.00325}.

\bibitem[{Cucerzan(2007)}]{cucerzan-2007-large}
Cucerzan, S. 2007.
\newblock Large-Scale Named Entity Disambiguation Based on {W}ikipedia Data.
\newblock In \emph{Proceedings of the 2007 Joint Conference on Empirical Methods in Natural Language Processing and Computational Natural Language Learning ({EMNLP}-{C}o{NLL})}, 708--716. Prague, Czech Republic: Association for Computational Linguistics.

\bibitem[{De~Cao et~al.(2020)De~Cao, Izacard, Riedel, and Petroni}]{de2020autoregressive}
De~Cao, N.; Izacard, G.; Riedel, S.; and Petroni, F. 2020.
\newblock Autoregressive entity retrieval.
\newblock \emph{arXiv preprint arXiv:2010.00904}.

\bibitem[{Gan et~al.(2021)Gan, Luo, Wang, Wang, He, and Huang}]{gan2021multimodal}
Gan, J.; Luo, J.; Wang, H.; Wang, S.; He, W.; and Huang, Q. 2021.
\newblock Multimodal Entity Linking: a New Dataset and a Baseline.
\newblock \emph{Multimedia}.

\bibitem[{Guo and Barbosa(2018)}]{guo2018robust}
Guo, Z.; and Barbosa, D. 2018.
\newblock Robust named entity disambiguation with random walks.
\newblock \emph{Semantic Web}, 9(4): 459--479.

\bibitem[{Hoffart et~al.(2011)Hoffart, Yosef, Bordino, F{\"u}rstenau, Pinkal, Spaniol, Taneva, Thater, and Weikum}]{hoffart-etal-2011-robust}
Hoffart, J.; Yosef, M.~A.; Bordino, I.; F{\"u}rstenau, H.; Pinkal, M.; Spaniol, M.; Taneva, B.; Thater, S.; and Weikum, G. 2011.
\newblock Robust Disambiguation of Named Entities in Text.
\newblock In \emph{Proceedings of the 2011 Conference on Empirical Methods in Natural Language Processing}, 782--792. Edinburgh, Scotland, UK.: Association for Computational Linguistics.

\bibitem[{Jia et~al.(2021)Jia, Yang, Xia, Chen, Parekh, Pham, Le, Sung, Li, and Duerig}]{align}
Jia, C.; Yang, Y.; Xia, Y.; Chen, Y.-T.; Parekh, Z.; Pham, H.; Le, Q.; Sung, Y.-H.; Li, Z.; and Duerig, T. 2021.
\newblock Scaling up visual and vision-language representation learning with noisy text supervision.
\newblock In \emph{International conference on machine learning}, 4904--4916. PMLR.

\bibitem[{Li et~al.(2023)Li, Li, Savarese, and Hoi}]{blip2}
Li, J.; Li, D.; Savarese, S.; and Hoi, S. 2023.
\newblock Blip-2: Bootstrapping language-image pre-training with frozen image encoders and large language models.
\newblock In \emph{International conference on machine learning}, 19730--19742. PMLR.

\bibitem[{Liu et~al.(2020)Liu, Wang, Wang, and Ordonez}]{liu2020visual}
Liu, F.; Wang, Y.; Wang, T.; and Ordonez, V. 2020.
\newblock Visual news: Benchmark and challenges in news image captioning.
\newblock \emph{arXiv preprint arXiv:2010.03743}.

\bibitem[{Logeswaran et~al.(2019)Logeswaran, Chang, Lee, Toutanova, Devlin, and Lee}]{logeswaran-etal-2019-zero}
Logeswaran, L.; Chang, M.-W.; Lee, K.; Toutanova, K.; Devlin, J.; and Lee, H. 2019.
\newblock Zero-Shot Entity Linking by Reading Entity Descriptions.
\newblock In \emph{Proceedings of the 57th Annual Meeting of the Association for Computational Linguistics}, 3449--3460. Florence, Italy: Association for Computational Linguistics.

\bibitem[{Lu et~al.(2018)Lu, Neves, Carvalho, Zhang, and Ji}]{lu2018visual}
Lu, D.; Neves, L.; Carvalho, V.; Zhang, N.; and Ji, H. 2018.
\newblock Visual attention model for name tagging in multimodal social media.
\newblock In \emph{Proceedings of the 56th Annual Meeting of the Association for Computational Linguistics (Volume 1: Long Papers)}, 1990--1999.

\bibitem[{Martinez-Rodriguez, Hogan, and Lopez-Arevalo(2020)}]{martinez2020information}
Martinez-Rodriguez, J.~L.; Hogan, A.; and Lopez-Arevalo, I. 2020.
\newblock Information extraction meets the semantic web: a survey.
\newblock \emph{Semantic Web}, 11(2): 255--335.

\bibitem[{Milne and Witten(2008)}]{milne2008learning}
Milne, D.; and Witten, I.~H. 2008.
\newblock Learning to link with wikipedia.
\newblock In \emph{Proceedings of the 17th ACM conference on Information and knowledge management}, 509--518.

\bibitem[{Minderer et~al.(2022)Minderer, Gritsenko, Stone, Neumann, Weissenborn, Dosovitskiy, Mahendran, Arnab, Dehghani, Shen, Wang, Zhai, Kipf, and Houlsby}]{OWL-ViT}
Minderer, M.; Gritsenko, A.; Stone, A.; Neumann, M.; Weissenborn, D.; Dosovitskiy, A.; Mahendran, A.; Arnab, A.; Dehghani, M.; Shen, Z.; Wang, X.; Zhai, X.; Kipf, T.; and Houlsby, N. 2022.
\newblock Simple Open-Vocabulary Object Detection with Vision Transformers.
\newblock arXiv:2205.06230.

\bibitem[{Moll{\'a}, Van~Zaanen, and Smith(2006)}]{molla2006named}
Moll{\'a}, D.; Van~Zaanen, M.; and Smith, D. 2006.
\newblock Named entity recognition for question answering.
\newblock In \emph{Australasian Language Technology Association Workshop}, 51--58. Australasian Language Technology Association.

\bibitem[{Moon, Neves, and Carvalho(2018)}]{moon-etal-2018-multimodal-named}
Moon, S.; Neves, L.; and Carvalho, V. 2018.
\newblock Multimodal Named Entity Disambiguation for Noisy Social Media Posts.
\newblock In \emph{Proceedings of the 56th Annual Meeting of the Association for Computational Linguistics (Volume 1: Long Papers)}, 2000--2008. Melbourne, Australia: Association for Computational Linguistics.

\bibitem[{Radford et~al.(2021)Radford, Kim, Hallacy, Ramesh, Goh, Agarwal, Sastry, Askell, Mishkin, Clark et~al.}]{clip}
Radford, A.; Kim, J.~W.; Hallacy, C.; Ramesh, A.; Goh, G.; Agarwal, S.; Sastry, G.; Askell, A.; Mishkin, P.; Clark, J.; et~al. 2021.
\newblock Learning transferable visual models from natural language supervision.
\newblock In \emph{International conference on machine learning}, 8748--8763. PMLR.

\bibitem[{Rao, McNamee, and Dredze(2013)}]{rao2013entity}
Rao, D.; McNamee, P.; and Dredze, M. 2013.
\newblock Entity linking: Finding extracted entities in a knowledge base.
\newblock \emph{Multi-source, multilingual information extraction and summarization}, 93--115.

\bibitem[{Ratinov et~al.(2011)Ratinov, Roth, Downey, and Anderson}]{ratinov-etal-2011-local}
Ratinov, L.; Roth, D.; Downey, D.; and Anderson, M. 2011.
\newblock Local and Global Algorithms for Disambiguation to {W}ikipedia.
\newblock In \emph{Proceedings of the 49th Annual Meeting of the Association for Computational Linguistics: Human Language Technologies}, 1375--1384. Portland, Oregon, USA: Association for Computational Linguistics.

\bibitem[{Sevgili et~al.(2022)Sevgili, Shelmanov, Arkhipov, Panchenko, and Biemann}]{sevgili2022neural}
Sevgili, {\"O}.; Shelmanov, A.; Arkhipov, M.; Panchenko, A.; and Biemann, C. 2022.
\newblock Neural entity linking: A survey of models based on deep learning.
\newblock \emph{Semantic Web}, 13(3): 527--570.

\bibitem[{Singh et~al.(2022)Singh, Hu, Goswami, Couairon, Galuba, Rohrbach, and Kiela}]{flava}
Singh, A.; Hu, R.; Goswami, V.; Couairon, G.; Galuba, W.; Rohrbach, M.; and Kiela, D. 2022.
\newblock Flava: A foundational language and vision alignment model.
\newblock In \emph{Proceedings of the IEEE/CVF Conference on Computer Vision and Pattern Recognition}, 15638--15650.

\bibitem[{Sun et~al.(2022)Sun, Fan, Guo, Zhang, and Cheng}]{sun2022visual}
Sun, W.; Fan, Y.; Guo, J.; Zhang, R.; and Cheng, X. 2022.
\newblock Visual named entity linking: A new dataset and a baseline.
\newblock \emph{arXiv preprint arXiv:2211.04872}.

\bibitem[{Tran, Mathews, and Xie(2020)}]{tran2020transform}
Tran, A.; Mathews, A.; and Xie, L. 2020.
\newblock Transform and tell: Entity-aware news image captioning.
\newblock In \emph{Proceedings of the IEEE/CVF conference on computer vision and pattern recognition}, 13035--13045.

\bibitem[{Wang et~al.(2022{\natexlab{a}})Wang, Tian, Gui, Li, Wang, Yan, Chen, and Xiao}]{wang2022wikidiverse}
Wang, X.; Tian, J.; Gui, M.; Li, Z.; Wang, R.; Yan, M.; Chen, L.; and Xiao, Y. 2022{\natexlab{a}}.
\newblock WikiDiverse: a multimodal entity linking dataset with diversified contextual topics and entity types.
\newblock \emph{arXiv preprint arXiv:2204.06347}.

\bibitem[{Wang et~al.(2022{\natexlab{b}})Wang, Montoya, Munechika, Yang, Hoover, and Chau}]{wangDiffusionDBLargescalePrompt2022}
Wang, Z.~J.; Montoya, E.; Munechika, D.; Yang, H.; Hoover, B.; and Chau, D.~H. 2022{\natexlab{b}}.
\newblock {{DiffusionDB}}: {{A}} Large-Scale Prompt Gallery Dataset for Text-to-Image Generative Models.
\newblock \emph{arXiv:2210.14896 [cs]}.

\bibitem[{Wu et~al.(2019)Wu, Petroni, Josifoski, Riedel, and Zettlemoyer}]{wu2019scalable}
Wu, L.; Petroni, F.; Josifoski, M.; Riedel, S.; and Zettlemoyer, L. 2019.
\newblock Scalable zero-shot entity linking with dense entity retrieval.
\newblock \emph{arXiv preprint arXiv:1911.03814}.

\bibitem[{Young et~al.(2014)Young, Lai, Hodosh, and Hockenmaier}]{young2014image}
Young, P.; Lai, A.; Hodosh, M.; and Hockenmaier, J. 2014.
\newblock From image descriptions to visual denotations: New similarity metrics for semantic inference over event descriptions.
\newblock \emph{Transactions of the Association for Computational Linguistics}, 2: 67--78.

\bibitem[{Zhai et~al.(2023)Zhai, Mustafa, Kolesnikov, and Beyer}]{siglip}
Zhai, X.; Mustafa, B.; Kolesnikov, A.; and Beyer, L. 2023.
\newblock Sigmoid loss for language image pre-training.
\newblock In \emph{Proceedings of the IEEE/CVF International Conference on Computer Vision}, 11975--11986.

\bibitem[{Zhang, Li, and Yang(2021)}]{zhang2021attention}
Zhang, L.; Li, Z.; and Yang, Q. 2021.
\newblock Attention-Based Multimodal Entity Linking with High-Quality Images.
\newblock In \emph{International Conference on Database Systems for Advanced Applications}, 533--548. Springer.

\bibitem[{Zhang et~al.(2018)Zhang, Fu, Liu, and Huang}]{zhang2018adaptive}
Zhang, Q.; Fu, J.; Liu, X.; and Huang, X. 2018.
\newblock Adaptive co-attention network for named entity recognition in tweets.
\newblock In \emph{Proceedings of the AAAI conference on artificial intelligence}, volume~32.

\end{thebibliography}


\end{document}